# An approach for analyzing the ensemble mean from a dynamic point of view

Wang Pengfei[1,2]

[1]Center for Monsoon System Research, Institute of Atmospheric Physics, Chinese Academy of Sciences, Beijing, 100190, China

[2]State Key Laboratory of Numerical Modeling for Atmospheric Sciences and Geophysical Fluid Dynamics (LASG), Institute of Atmospheric Physics, Chinese Academy of Sciences, Beijing, 100029, China

**Abstract:** Simultaneous ensemble mean equations (LEMEs) for the Lorenz model are obtained, enabling us to analyze the properties of the ensemble mean from a dynamical point of view. The qualitative analysis for the two-sample and *n*-sample LEMEs show the locations and number of stable points are different from the Lorenz equations (LEs), and the results are validated by numerical experiments. The analysis for the eigenmatrix of the stable points of LEMEs indicates that the stability of these stable points is similar to the LEs'. The eigenmatrix for non-stable points can be obtained too, but the eigenvalues depend not only on the value of the mean variable but also the other $n-1$ sample equation's variable, and thus for these points there may be different stabilities compared to the LEs'. The divergence of the LEMEs' flow has a negative value, which is the same as the LEs', and thus the trajectory in phase space approaches zero and the trajectory will be attracted to a low-level dimensional curved surface, i.e., the LEMEs have the attractor property, but the structure of the attractor is not the same as the LEs'.

## 1 Introduction

Many researchers have contributed to the theory and application of ensemble prediction [1-9]. Ensemble forecasts are widely used in short-term weather forecasting as well as in the prediction of monthly, seasonal, annual and decade climate. The aim of ensemble forecasting is to improve the forecast quality and provide a quantitative forecast product. The ensemble mean operation is one of the simplest and most commonly used methods. Previous studies have generally focused on the ensemble forecast from the stochastic point of view and applied it in the prediction of weather and climate. However, more recently, some researchers are now also interested in studying whether or not the ensemble method can reduce the error in the prediction of chaotic systems. Houtekamer [10, 11] used the Lorenz63 equation to study a two-sample ensemble forecast, and his experiments indicated that the difference between the control experiment and the results of two initial perturbation ensemble samples was due to nonlinear processes, and thus a nonlinear model is the necessary requirement for studying ensembles. Anderson [12] studied constrained and unconstrained ensembles in a nonlinear system by two low-order models, and the results indicated that unconstrained ensembles can achieve the equivalent outcome as constrained ones.

Qualitative theory of ordinary equations has many successful applications in investigating the trajectories and chaotic attractors of dynamical systems. In this study, we regard the ensemble mean as an initial value problem of the dynamic system with ensemble operation, and apply qualitative theory to study its dynamical properties and attractor structure. This will help to increase our knowledge and improve our understanding of ensemble characteristics.

The remainder of the paper is organized as follows. In section 2, we analyze the dynamic behavior of a two-samples ensemble mean for a Lorenz system, and apply stability theory to calculate the fixed points. In section 3, the stabilities of the fixed points for the two-sample ensemble mean are obtained and proved to be the same as the original dynamical system. In sections 4 and 5, we extend the two-sample results to an *n*-sample ensemble mean and a general dynamical system. And finally, section 6 provides a summary of the study.

## 2 Stability analysis for a two-sample ensemble mean of a Lorenz system

Lorenz found that solutions of a deterministic nonlinear dynamical system may be very complicated [13]. The Lorenz63 equation is widely used in chaos research and predictability studies and can be expressed as:

$$\begin{cases} \dfrac{dx}{dt} = -\sigma x + \sigma y \\ \dfrac{dy}{dt} = rx - y - xz, \\ \dfrac{dz}{dt} = xy - bz \end{cases} \tag{1}$$

where $\sigma$ , $r$ , and $b$ are constants (e.g., $\sigma = 10, r = 28.0, b = 8/3$ ) , and $t$ is non-dimensional time. We denote the reference as $(x, y, z)$, the initial value for the reference solution as $(x^0, y^0, z^0) = (5,10,5)$, and we gather a two-sample ensemble mean: the first case has an initial value of $(x_1^0, y_1^0, z_1^0) = (5.01,10,5)$; the second case has an initial value of $(x_2^0, y_2^0, z_2^0) = (4.99,10,5)$; the mean value at time $t$ is $(\overline{x}, \overline{y}, \overline{z})$; and the equations for the mean variables are:

$$\begin{cases} \dfrac{dx_1}{dt} = -\sigma x_1 + \sigma y_1 \\ \dfrac{dy_1}{dt} = rx_1 - y_1 - x_1 z_1, \\ \dfrac{dz_1}{dt} = x_1 y_1 - bz_1 \end{cases} \tag{2}$$

$$\begin{cases} \dfrac{dx_2}{dt} = -\sigma x_2 + \sigma y_2 \\ \dfrac{dy_2}{dt} = rx_2 - y_2 - x_2 z_2, \\ \dfrac{dz_2}{dt} = x_2 y_2 - bz_2 \end{cases} \tag{3}$$

Since $\overline{x} = \frac{1}{2}(x_1 + x_2), \overline{y} = \frac{1}{2}(y_1 + y_2), \overline{z} = \frac{1}{2}(z_1 + z_2)$, thus the initial value for the mean variable is $\frac{1}{2}(x_1^0 + x_2^0), \frac{1}{2}(y_1^0 + y_2^0), \frac{1}{2}(z_1^0 + z_2^0)$.

Accord to (2) and (3) we can eliminate $(x_2, y_2, z_2)$ to obtain Eq. (4):

$$\begin{cases} \frac{dx_1}{dt} = -\sigma x_1 + \sigma y_1 \\ \frac{dy_1}{dt} = rx_1 - y_1 - x_1 z_1 \\ \frac{dz_1}{dt} = x_1 y_1 - bz_1 \\ \frac{d\overline{x}}{dt} = -\sigma \overline{x} + \sigma \overline{y} \\ \frac{d\overline{y}}{dt} = r\overline{x} - \overline{y} - \frac{1}{2}\left(2\overline{x} - x_1\right)\left(2\overline{z} - z_1\right) - \frac{1}{2} x_1 z_1 \\ \frac{d\overline{z}}{dt} = \frac{1}{2}\left(2\overline{x} - x_1\right)\left(2\overline{y} - y_1\right) + \frac{1}{2} x_1 y_1 - b\overline{z} \end{cases}, \tag{4}$$

where we have six variables and six equations. We can obtain the solutions of $\left(x_1, y_1, z_1, \overline{x}, \overline{y}, \overline{z}\right)$, and Eq. (4) describes the whole law to which the ensemble mean abides.

The stable point (or fixed point) for Eq. (4) can be solved by the equations

$$\begin{cases} 0 = -\sigma x_1 + \sigma y_1 \\ 0 = rx_1 - y_1 - x_1 z_1 \\ 0 = x_1 y_1 - bz_1 \\ 0 = -\sigma \overline{x} + \sigma \overline{y} \\ 0 = r\overline{x} - \overline{y} - \tfrac{1}{2}\left(2\overline{x} - x_1\right)\left(2\overline{z} - z_1\right) - \tfrac{1}{2} x_1 z_1 \\ 0 = \tfrac{1}{2}\left(2\overline{x} - x_1\right)\left(2\overline{y} - y_1\right) + \tfrac{1}{2} x_1 y_1 - b\overline{z} \end{cases},$$

which is a special equation because the first three equations are independent of the last three equations, but the solutions of the last three equations are affected by the first three. We can obtain the solutions of the first three equations and then solve the last three equations. When the solution is $x_1 = 0, y_1 = 0, z_1 = 0$, the last three equations have three sets of solutions: $\begin{cases} \overline{x} = 0 \\ \overline{y} = 0 \\ \overline{z} = 0 \end{cases}$; $\begin{cases} \overline{x} = \frac{1}{2}\sqrt{b\left(r-1\right)} \\ \overline{y} = \frac{1}{2}\sqrt{b\left(r-1\right)} \\ \overline{z} = \frac{1}{2}\left(r-1\right) \end{cases}$; $\begin{cases} \overline{x} = -\tfrac{1}{2}\sqrt{b\left(r-1\right)} \\ \overline{y} = -\tfrac{1}{2}\sqrt{b\left(r-1\right)} \\ \overline{z} = \tfrac{1}{2}\left(r-1\right) \end{cases}$. Further, when $\begin{cases} x_1 = \sqrt{b\left(r-1\right)} \\ y_1 = \sqrt{b\left(r-1\right)} \\ z_1 = r-1 \end{cases}$, another three solutions are: $\begin{cases} \overline{x} = 0 \\ \overline{y} = 0 \\ \overline{z} = r-1 \end{cases}$; $\begin{cases} \overline{x} = \tfrac{1}{2}\sqrt{b\left(r-1\right)} \\ \overline{y} = \tfrac{1}{2}\sqrt{b\left(r-1\right)} \\ \overline{z} = \tfrac{1}{2}\left(r-1\right) \end{cases}$ (duplicated solution); $\begin{cases} \overline{x} = \sqrt{b\left(r-1\right)} \\ \overline{y} = \sqrt{b\left(r-1\right)} \\ \overline{z} = \left(r-1\right) \end{cases}$.

Finally, when $\begin{cases} x_1 = -\sqrt{b(r-1)} \\ y_1 = -\sqrt{b(r-1)} \\ z_1 = r-1 \end{cases}$, the three solutions are: $\begin{cases} \bar{x} = 0 \\ \bar{y} = 0 \\ \bar{z} = r-1 \end{cases}$ (duplicated solution); $\begin{cases} \bar{x} = -\frac{1}{2}\sqrt{b(r-1)} \\ \bar{y} = -\frac{1}{2}\sqrt{b(r-1)} \\ \bar{z} = \frac{1}{2}(r-1) \end{cases}$ (duplicated solution); $\begin{cases} \bar{x} = -\sqrt{b(r-1)} \\ \bar{y} = -\sqrt{b(r-1)} \\ \bar{z} = (r-1) \end{cases}$.

Discarding the location of $x_1, y_1, z_1$ and only considering the center of $(\bar{x}, \bar{y}, \bar{z})$, we find that each mean variable will have six center points (dropping the three duplicated solutions among the nine solutions), which are: $\begin{cases} \bar{x} = 0 \\ \bar{y} = 0 \\ \bar{z} = 0 \end{cases}$; $\begin{cases} \bar{x} = \frac{1}{2}\sqrt{b(r-1)} \\ \bar{y} = \frac{1}{2}\sqrt{b(r-1)} \\ \bar{z} = \frac{1}{2}(r-1) \end{cases}$; $\begin{cases} \bar{x} = -\frac{1}{2}\sqrt{b(r-1)} \\ \bar{y} = -\frac{1}{2}\sqrt{b(r-1)} \\ \bar{z} = \frac{1}{2}(r-1) \end{cases}$; $\begin{cases} \bar{x} = 0 \\ \bar{y} = 0 \\ \bar{z} = r-1 \end{cases}$; $\begin{cases} \bar{x} = \sqrt{b(r-1)} \\ \bar{y} = \sqrt{b(r-1)} \\ \bar{z} = (r-1) \end{cases}$; $\begin{cases} \bar{x} = -\sqrt{b(r-1)} \\ \bar{y} = -\sqrt{b(r-1)} \\ \bar{z} = (r-1) \end{cases}$.

We can investigate the two-dimensional phase structure of $\bar{x}, \bar{y}$ and neglect the variation of $\bar{z}$, and we find five centers in the image ($\begin{cases} \bar{x} = 0 \\ \bar{y} = 0 \\ \bar{z} = 0 \end{cases}$ and $\begin{cases} \bar{x} = 0 \\ \bar{y} = 0 \\ \bar{z} = r-1 \end{cases}$ are mixed together when projected on the plane). This is quite different to the original dynamical system, which only has three centers.

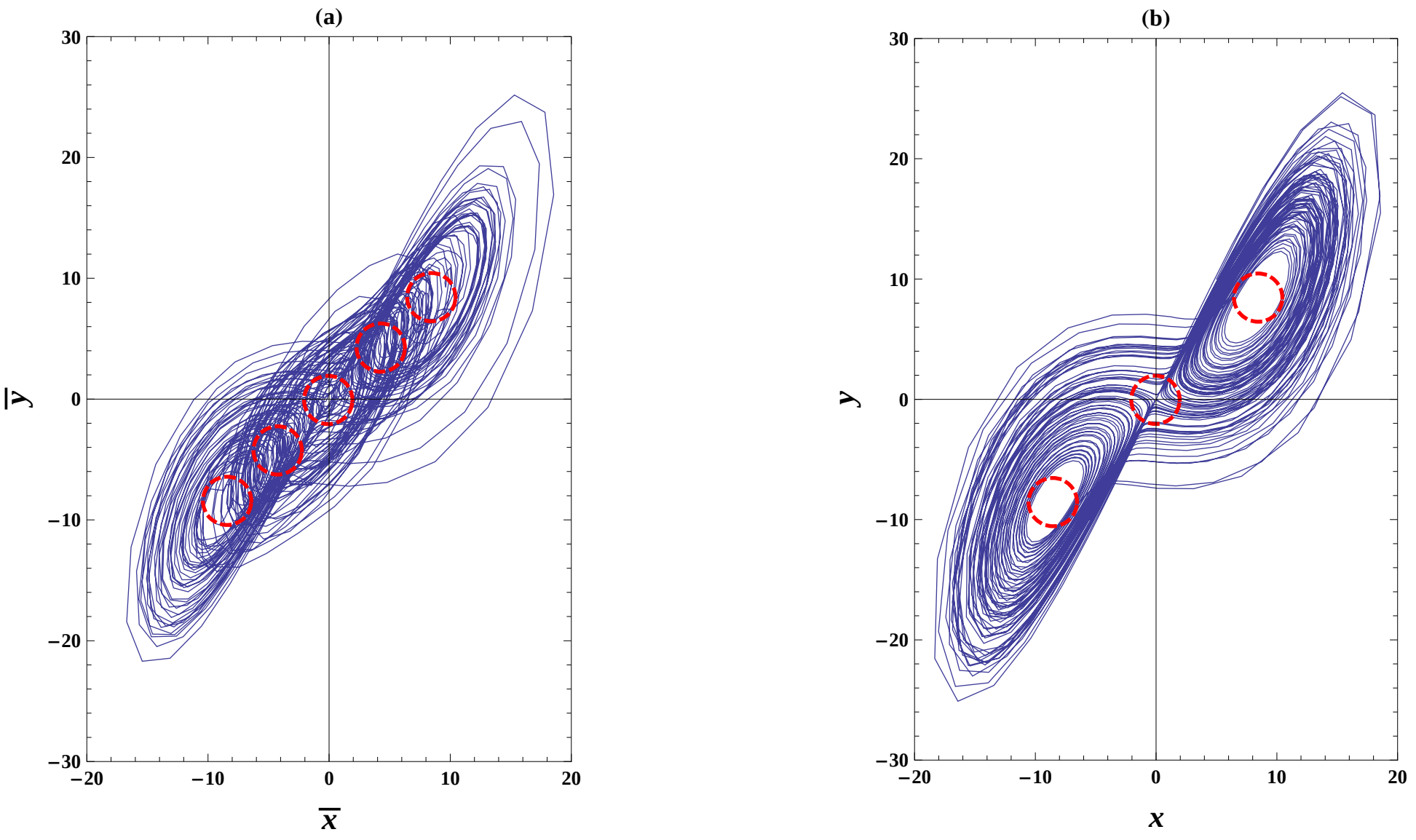

(a) (b)

**Fig. 1.** Solution of Eq. (4) for time $t = 0 - 100$ with the initial condition $\left(x_1^0, y_1^0, z_1^0\right) = \left(5.01, 10, 5\right)$ and $\left(\overline{x^0}, \overline{y^0}, \overline{z^0}\right) = \left(5, 10, 5\right)$. The dashed red circle indicates the location of the stable points. (a) The projection of error ($\overline{x}, \overline{y}, \overline{z}$) on the ($\overline{x}, \overline{y}$) plane; (b) the projection of the variable ($x, y, z$) on the ($x, y$) plane.

Figure 1a demonstrates via the numerical solution of the mean variables' ($\overline{x}, \overline{y}, \overline{z}$) projection onto the ($\overline{x}, \overline{y}$) plane that the trajectory is complex, and the five center locations are marked with dashed circles. Figure 1b is the projection of variable $(x, y, z)$ onto the $(x, y)$ plane, which is different from Fig. 1a: there are three center points in Fig. 1b. The mean variables have different numbers and locations to the original variables.

## 3 The stability of fixed points for the two-sample ensemble mean Lorenz system

In total, Eq. (4) has nine stable points, and we can analyze the stability at these points through qualitative theory. For example, we perform perturbation analysis at the stable points for Eq. (4) and omit the high-order items:

$$
\begin{cases}
\dfrac{dx_1'}{dt} = -\sigma x_1' + \sigma y_1' \\
\dfrac{dy_1'}{dt} = r x_1' - y_1' - x_1' z_1 - x_1 z_1' \\
\dfrac{dz_1'}{dt} = x_1' y_1 + x_1 y_1' - b z_1' \\
\dfrac{d\bar{x}'}{dt} = -\sigma \bar{x}' + \sigma \bar{y}' \\
\dfrac{d\bar{y}'}{dt} = r\bar{x}' - \bar{y}' - 2\overline{xz}' - 2\bar{x}'\bar{z} + x_1'\bar{z} + x_1\bar{z}' + \bar{x}' z_1 + \bar{x} z_1' - x_1' z_1 - x_1 z_1' \\
\dfrac{d\bar{z}'}{dt} = 2\overline{xy}' + 2\bar{x}'\bar{y} - x_1'\bar{y} - x_1\bar{y}' - \bar{x}' y_1 - \bar{x} y_1' + x_1' y_1 + x_1 y_1' - b\bar{z}'
\end{cases}
. \tag{5}
$$

The linearization matrix of (5) is:

$$
\begin{bmatrix}
-\sigma & \sigma & 0 & 0 & 0 & 0 \\
(r - z_1) & -1 & -x_1 & 0 & 0 & 0 \\
y_1 & x_1 & -b & 0 & 0 & 0 \\
0 & 0 & 0 & -\sigma & \sigma & 0 \\
\bar{z} - z_1 & 0 & \bar{x} - x_1 & (r + z_1 - 2\bar{z}) & -1 & (-2\bar{x} + x_1) \\
y_1 - \bar{y} & x_1 - \bar{x} & 0 & (2\bar{y} - y_1) & (2\bar{x} - x_1) & -b
\end{bmatrix}
. \tag{6}
$$

The matrix (6) can be used to investigate the stability around a fixed point. When $x_1 = 0, y_1 = 0, z_1 = 0, \bar{x} = 0, \bar{y} = 0, \bar{z} = 0$, the linearization matrix (6) is

$$
\begin{bmatrix}
-\sigma & \sigma & 0 & 0 & 0 & 0 \\
r & -1 & 0 & 0 & 0 & 0 \\
0 & 0 & -b & 0 & 0 & 0 \\
0 & 0 & 0 & -\sigma & \sigma & 0 \\
0 & 0 & 0 & r & -1 & 0 \\
0 & 0 & 0 & 0 & 0 & -b
\end{bmatrix}
.
$$

Suppose the eigenvalue is $\lambda$, then the eigen equation is $(\lambda + b)^2 \left[\lambda^2 + (\sigma + 1)\lambda + \sigma(1 - r)\right]^2 = 0$.

The roots of the eigen equation are:

$$\lambda_1 = -b, \lambda_{2,3} = \frac{-(\sigma+1) \pm \sqrt{(\sigma+1)^2 + 4\sigma(r-1)}}{2};$$

$$\lambda_4 = -b, \lambda_{5,6} = \frac{-(\sigma+1) \pm \sqrt{(\sigma+1)^2 + 4\sigma(r-1)}}{2}.$$

If $r < 1$, the fixed point is a stable attractive point; if $r = 1$, the fixed point is a separate point; and when $r > 1$, the fixed point becomes an unstable saddle point.

The second sets of solutions, $\begin{cases} x_1 = 0 \\ y_1 = 0 \\ z_1 = 0 \end{cases}$, with $\begin{cases} \bar{x} = ½\sqrt{b(r-1)} \\ \bar{y} = ½\sqrt{b(r-1)} \\ \bar{z} = ½(r-1) \end{cases}$, means Eq. (6) becomes

$$\begin{bmatrix} -\sigma & \sigma & 0 & 0 & 0 & 0 \\ r & -1 & 0 & 0 & 0 & 0 \\ 0 & 0 & -b & 0 & 0 & 0 \\ 0 & 0 & 0 & -\sigma & \sigma & 0 \\ (r-1)/2 & 0 & \sqrt{b(r-1)}/2 & 1 & -1 & -\sqrt{b(r-1)} \\ -\sqrt{b(r-1)}/2 & -\sqrt{b(r-1)}/2 & 0 & \sqrt{b(r-1)} & \sqrt{b(r-1)} & -b \end{bmatrix}, \tag{7}$$

and the eigen equation of (7) is

$$\left[\lambda^3 + (\sigma+b+1)\lambda^2 + (r+\sigma)b\lambda + 2\sigma b(r-1)\right](\lambda+b)\left[\lambda^2 + (\sigma+1)\lambda + \sigma(1-r)\right] = 0. \tag{8}$$

The third set of solutions is $\begin{cases} x_1 = 0 \\ y_1 = 0 \\ z_1 = 0 \end{cases}$, with $\begin{cases} \bar{x} = -\sqrt{b(r-1)}/2 \\ \bar{y} = -\sqrt{b(r-1)}/2 \\ \bar{z} = (r-1)/2 \end{cases}$, meaning Eq. (6) becomes

$$\begin{bmatrix} -\sigma & \sigma & 0 & 0 & 0 & 0 \\ r & -1 & 0 & 0 & 0 & 0 \\ 0 & 0 & -b & 0 & 0 & 0 \\ 0 & 0 & 0 & -\sigma & \sigma & 0 \\ (r-1)/2 & 0 & -\sqrt{b(r-1)}/2 & 1 & -1 & \sqrt{b(r-1)} \\ \sqrt{b(r-1)}/2 & \sqrt{b(r-1)}/2 & 0 & -\sqrt{b(r-1)} & -\sqrt{b(r-1)} & -b \end{bmatrix}, \tag{9}$$

and the eigen equation of (9) is

$$\left[\lambda^3+(\sigma+b+1)\lambda^2+(r+\sigma)b\lambda+2\sigma b(r-1)\right](\lambda+b)\left[\lambda^2+(\sigma+1)\lambda+\sigma(1-r)\right]=0\,. \quad (10)$$

We repeat the same procedure for the other six sets of solutions and find that, despite the linearization matrix being six-order, the eigenvalues at the fixed points are the same as the eiginvalues at the fixed points in the original Lorenz system. In the application of stability analysis, the positive property of the real part in the complex root determines if the system is stable or divergent, and thus the range analysis of $r$ for the Lorenz system is also suitable for this ensemble mean variable equation; that is, when $1<r<24.74$, the system is stable, and when $r>24.74$ the system becomes chaotic.

In fact, the eignematrix (6) and its corresponding eigenvalues determine if the local trajectory is divergent or convergent. We can easily establish that these eigenvalues not only depend on the location of the mean variable, but also the other variable's location, which is involved in the ensemble mean. This is the main difference between the ensemble mean dynamical system and the original Lorenz system.

## 4 The *n*-sample ensemble mean of a Lorenz system

We extend our analysis from a two-sample to $n$-sample ensemble. Each initial condition has a solution following the equation:

$$\begin{cases}\dfrac{dx_i}{dt}=-\sigma x_i+\sigma y_i\\ \dfrac{dy_i}{dt}=rx_i-y_i-x_iz_i\\ \dfrac{dz_i}{dt}=x_iy_i-bz_i\end{cases}, \quad (11)$$

where $i$ is an integer and $1\le i\le n$, and the initial values are $\left(x_i^0,y_i^0,z_i^0\right)$. We perform variable transformation to the last three equations of (11) as $x_n=n\bar{x}-\sum_{i=1}^{n-1}x_i, y_n=n\bar{y}-\sum_{i=1}^{n-1}y_i, z_n=n\bar{z}-\sum_{i=1}^{n-1}z_i$, and then Eq. (11) is converted to an equation formulate by $x_1,x_2,\cdots,x_{n-1},\bar{x}$, $y_1,y_2,\cdots,y_{n-1},\bar{y}$, $z_1,z_2,\cdots,z_{n-1},\bar{z}$. The fixed point of this ensemble mean system is $\begin{cases} \bar{x} = \frac{1}{n}\sum_{i=1}^{n} X_i^k \\ \bar{y} = \frac{1}{n}\sum_{i=1}^{n} Y_i^k \\ \bar{z} = \frac{1}{n}\sum_{i=1}^{n} Z_i^k \end{cases}$, where $X_i^k, Y_i^k, Z_i^k$ is the $k$-th fixed point of Eq. (1), and $1 \le k \le 3$. This formula tells us the fixed point of the ensemble mean system is the sum of the different arrange of (1), and then divided by $n$, there will mostly be $3^n$ fixed points (including duplicated ones). A more detailed proof and discussion can be found in section 5.

The last three equations of (11) are translated to the new format:

$$\begin{cases} \frac{d\bar{x}}{dt} = -\sigma\bar{x} + \sigma\bar{y} \\ \frac{d\bar{y}}{dt} = r\bar{x} - \bar{y} - \frac{1}{n}\left(n\bar{x} - \sum_{i=1}^{n-1} x_i\right)\left(n\bar{z} - \sum_{i=1}^{n-1} z_i\right) - \frac{1}{n}\sum_{i=1}^{n-1} x_i z_i \\ \frac{d\bar{z}}{dt} = \frac{1}{n}\left(n\bar{x} - \sum_{i=1}^{n-1} x_i\right)\left(n\bar{y} - \sum_{i=1}^{n-1} y_i\right) + \frac{1}{n}\sum_{i=1}^{n-1} x_i y_i - b\bar{z} \end{cases} . \tag{12}$$

When we take the divergence operation in Eq. (11), we obtain the divergence value $-n(\sigma + 1 + b)$. Thus the divergence of the mean variable dynamic is negative and the volume of this system in phase space is smaller and smaller. This property is the same as that possessed by the Lorenz system. In addition, we can prove that an attractor exists for the mean variable dynamics by applying the method Wang et al. [14]; the result indicates a global attractor property, which is also the same in the Lorenz system.

To summarize the investigation made in sections 2, 3 and 4, we know that the ensemble mean Lorenz system has different properties to the original in terms of their stability, the location of the fixed points, and the eigenvalues of the local Jacobian matrix; but the same properties of dissipative and attractor existence.

## 5 The *n*-sample ensemble mean of a general dynamical system

The $m$ variable ordinary differential system is denoted as

$$\begin{cases} \dfrac{dx_1}{dt} = F_1\left(x_1, x_2, \cdots, x_m\right) \\ \dfrac{dx_2}{dt} = F_2\left(x_1, x_2, \cdots, x_m\right) \\ \qquad\vdots \\ \dfrac{dx_m}{dt} = F_m\left(x_1, x_2, \cdots, x_m\right) \end{cases}, \tag{13}$$

where $F_j\left(x_1, x_2, \cdots, x_m\right)$ is a function. The fixed point problem of (13) is

$$\begin{cases} 0 = F_1\left(x_1, x_2, \cdots, x_m\right) \\ 0 = F_2\left(x_1, x_2, \cdots, x_m\right) \\ \qquad\vdots \\ 0 = F_m\left(x_1, x_2, \cdots, x_m\right) \end{cases}. \tag{14}$$

Suppose there are $K$ fixed points for Eq. (14), and the $k$-th fixed point is $X^k = \left(X_1^k, X_2^k, \cdots, X_m^k\right)$. Denote $x_{j,i}$ as the $i$-th solution for the $j$-th variable, and then the mean variable equation according to Eq. (13) is

$$
\begin{cases}
\frac{dx_{1,1}}{dt} = F_1\left(x_{1,1}, x_{2,1}, \cdots, x_{m,1}\right) \\
\frac{dx_{2,1}}{dt} = F_2\left(x_{1,1}, x_{2,1}, \cdots, x_{m,1}\right) \\
\vdots \\
\frac{dx_{m,1}}{dt} = F_m\left(x_{1,1}, x_{2,1}, \cdots, x_{m,1}\right) \\
\frac{dx_{1,2}}{dt} = F_1\left(x_{1,2}, x_{2,2}, \cdots, x_{m,2}\right) \\
\frac{dx_{2,2}}{dt} = F_2\left(x_{1,2}, x_{2,2}, \cdots, x_{m,2}\right) \\
\vdots \\
\frac{dx_{m,2}}{dt} = F_m\left(x_{1,2}, x_{2,2}, \cdots, x_{m,2}\right) \\
\frac{dx_{1,n}}{dt} = F_1\left(x_{1,n}, x_{2,n}, \cdots, x_{m,n}\right) \\
\frac{dx_{2,n}}{dt} = F_2\left(x_{1,n}, x_{2,n}, \cdots, x_{m,n}\right) \\
\vdots \\
\frac{dx_{m,n}}{dt} = F_m\left(x_{1,n}, x_{2,n}, \cdots, x_{m,n}\right)
\end{cases}
\tag{15}
$$

The last $m$ equations can be altered by $\left(\overline{x_1}, \overline{x_2}, \cdots, \overline{x_m}\right)$ for variable $\left(x_{1,n}, x_{2,n}, \cdots, x_{m,n}\right)$, and thus the fixed point problem of Eq. (15) is

$$
\begin{cases}
0 = F_1\left(x_{1,1}, x_{2,1}, \cdots, x_{m,1}\right) \\
0 = F_2\left(x_{1,1}, x_{2,1}, \cdots, x_{m,1}\right) \\
\vdots \\
0 = F_m\left(x_{1,1}, x_{2,1}, \cdots, x_{m,1}\right) \\
0 = F_1\left(x_{1,2}, x_{2,2}, \cdots, x_{m,2}\right) \\
0 = F_2\left(x_{1,2}, x_{2,2}, \cdots, x_{m,2}\right) \\
\vdots \\
0 = F_m\left(x_{1,2}, x_{2,2}, \cdots, x_{m,2}\right) \\
0 = F_1\left(x_{1,n}, x_{2,n}, \cdots, x_{m,n}\right) \\
0 = F_2\left(x_{1,n}, x_{2,n}, \cdots, x_{m,n}\right) \\
\vdots \\
0 = F_m\left(x_{1,n}, x_{2,n}, \cdots, x_{m,n}\right)
\end{cases}
\tag{16}
$$

We observe that each $m$ equation in (16) has the same structure to (14), and thus the fixed point number $\left(x_{1,i}, x_{2,i}, \cdots, x_{m,i}\right)$ is $K$, and their values are the same as the solution of (14).

Substituting the variable in the last $m$ equations in (16), i.e., $x_{j,n} = n\overline{x_j} - \sum_{i=1}^{n-1} x_{j,i}$, we obtain

$$\begin{cases} 0 = F_1\left(x_{1,n}, x_{2,n}, \cdots, x_{m,n}\right) \\ 0 = F_2\left(x_{1,n}, x_{2,n}, \cdots, x_{m,n}\right) \\ \qquad \vdots \\ 0 = F_m\left(x_{1,n}, x_{2,n}, \cdots, x_{m,n}\right) \end{cases}, \tag{17}$$

Equation (17) is the same as (14), and thus the fixed point number $\left(x_{1,n}, x_{2,n}, \cdots, x_{m,n}\right)$ is also $K$, and their values are the same as the solution of (14). The *k*-th fixed point in $\left(x_{1,n}, x_{2,n}, \cdots, x_{m,n}\right)$ satisfies the relation

$$\begin{cases} X_1^k = n\overline{x_1} - \sum_{i=1}^{n-1} x_{1,i} \\ X_2^k = n\overline{x_2} - \sum_{i=1}^{n-1} x_{2,i} \\ \vdots \\ X_m^k = n\overline{x_m} - \sum_{i=1}^{n-1} x_{m,i} \end{cases},$$

where $1 \le k \le K$, and $\left(x_{1,i}, x_{2,i}, \cdots, x_{m,i}\right)$ can be any choice from the $K$-th fixed point, and thus $\begin{cases} \overline{x_1} = \left(X_1^k + \sum_{i=1}^{n-1} x_{1,i}\right)\Big/n \\ \overline{x_2} = \left(X_2^k + \sum_{i=1}^{n-1} x_{2,i}\right)\Big/n \\ \vdots \\ \overline{x_m} = \left(X_m^k + \sum_{i=1}^{n-1} x_{m,i}\right)\Big/n \end{cases}$. This formula indicates that the fixed point of the ensemble system is the sum of the combinations of the fixed point to (14) and then, divided by $n$, there will mostly be $K^n$ fixed points (including duplications).

As we know the fixed point of the ensemble mean system according to the stability theory of a dynamical system, we only need to know the linearization matrix of (15) to decide the system's properties around the fixed point. This linearization matrix for Eq. (13) is denoted as

$$A=\begin{bmatrix} a_{11} & \cdots & a_{1m} \\ \vdots & \ddots & \vdots \\ a_{m1} & \cdots & a_{mm} \end{bmatrix}, \tag{18}$$

where $a_{lj}=\dfrac{\partial F_l}{\partial x_j}$, and for the mean variable system, Eq.(15), the linearization matrix is

$$C=\begin{bmatrix} a_{11,1} & \cdots & a_{1m,1} & 0 & \cdots & 0 & 0 & \cdots & 0 & 0 & \cdots & 0 \\ \vdots & \ddots & \vdots & \vdots & \ddots & \vdots & \vdots & \ddots & \vdots & \vdots & \ddots & \vdots \\ a_{m1,1} & \cdots & a_{mm,1} & 0 & \cdots & 0 & 0 & \cdots & 0 & 0 & \cdots & 0 \\ 0 & \cdots & 0 & a_{11,2} & \cdots & a_{1m,2} & 0 & \cdots & 0 & 0 & \cdots & 0 \\ \vdots & \ddots & \vdots & \vdots & \ddots & \vdots & \vdots & \ddots & \vdots & \vdots & \ddots & \vdots \\ 0 & \cdots & 0 & a_{m1,2} & \cdots & a_{mm,2} & 0 & \cdots & 0 & 0 & \cdots & 0 \\ 0 & \cdots & 0 & 0 & \cdots & 0 & a_{11,i} & \cdots & a_{1m,i} & 0 & \cdots & 0 \\ \vdots & \ddots & \vdots & \vdots & \ddots & \vdots & \vdots & \ddots & \vdots & \vdots & \ddots & \vdots \\ 0 & \cdots & 0 & 0 & \cdots & 0 & a_{m1,i} & \cdots & a_{mm,i} & 0 & \cdots & 0 \\ 0 & \cdots & 0 & 0 & \cdots & 0 & 0 & \cdots & 0 & a_{11,n} & \cdots & a_{1m,n} \\ \vdots & \ddots & \vdots & \vdots & \ddots & \vdots & \vdots & \ddots & \vdots & \vdots & \ddots & \vdots \\ 0 & \cdots & 0 & 0 & \cdots & 0 & 0 & \cdots & 0 & a_{m1,n} & \cdots & a_{mm,n} \end{bmatrix}, \tag{19}$$

where $a_{lj,i}=\dfrac{\partial F_l}{\partial x_j}$ and $A_i=\begin{bmatrix} a_{11,i} & \cdots & a_{1m,i} \\ \vdots & \ddots & \vdots \\ a_{m1,i} & \cdots & a_{mm,i} \end{bmatrix}$. The determinant value according to $A_i$ is

$$|A_i|=\begin{vmatrix} a_{11,i} & \cdots & a_{1m,i} \\ \vdots & \ddots & \vdots \\ a_{m1,i} & \cdots & a_{mm,i} \end{vmatrix}.$$

Linear algebraic knowledge tells us that the determinant value of (19) is $\det(C)=\prod_{i=1}^{n}\det(A_i)$. Substituting the variable with $x_{j,n}=n\overline{x_j}-\sum_{i=1}^{n-1}x_{j,i}$ and applying the relation $a_{lj,n} = \frac{\partial F_l}{\partial x_j}\left(x_{1,n}, x_{2,n}, \cdots, x_{m,n}\right)$, we can express $A_n = \begin{bmatrix} a_{11,n} & \cdots & a_{1m,n} \\ \vdots & \ddots & \vdots \\ a_{m1,n} & \cdots & a_{mm,n} \end{bmatrix}$ to a matrix which uses the $\overline{x_j}$ as the element, denoted as $B = \begin{bmatrix} b_{11} & \cdots & b_{1m} \\ \vdots & \ddots & \vdots \\ b_{m1} & \cdots & b_{mm} \end{bmatrix}$ (it should be noted that after the substitution of the variable, the last *m* rows of matrix $C$ have been changed, but the determinant value is only affected by the values in region $B$ ).

The eigenvalues according to (18) satisfy the equation $\det\left(\lambda I - A\right) = 0$, and for each fixed point we obtain a group of eigenvalues, $\lambda = \left(\lambda_1, \lambda_2, \cdots, \lambda_m\right)$, and in total there are $K$ groups. The eigenvalue equation of the mean variable system is $\det\left(\lambda I - B\right)\prod_{i=1}^{n-1}\left(\lambda I - A_i\right) = 0$, but $\det\left(\lambda I - B\right) = 0$ will obtain $m$ eigenvalues, which is only a rearrangement of the eigenvalues that satisfy $\det\left(\lambda I - A\right) = 0$. No new eigenvalues appear, and the maximal eigenvalue, $\lambda_{\max}$, is also the same. This result indicates that the parameters and their stability effects in (18) are also suitable for the ensemble mean variable system.

## 6 Conclusion

Qualitative analysis indicates that the two-sample ensemble mean equation of a Lorenz system has nine fixed point solutions, and if the location of the samples' fixed points are not considered, the mean variable is six. If the sample size changes from 2 to *n*, the theoretical number of fixed points is $3^n$, but there are many duplications within that number.

We then obtained the relation of the fixed points and their stabilities between the ensemble mean system and the original dynamical system in a general case by matrix analysis. The result suggests that, around the fixed points, there are no new eigenvalues for the ensemble mean system compared to the original system. Further, the eigenvalues are only a different sort of those that the original system has, and thus the stabilities and criteria are also suitable for the mean variable system. However, the stabilities around no-fixed points not only depend on the mean variable value, but also on the other sample's value, which participated in the ensemble operation.

The benefits of investigating the ensemble mean operation as a dynamical system is that we can apply all the qualitative analysis theories and methods that have been developed for the ordinary equations to investigate this ensemble mean system. This will help us to understand many of the ensemble mean characteristics. We have listed some properties that are not mentioned by traditional ensemble theory, such as fixed points, stability, and attractors of the ensemble mean system. This new concept can easily describe which properties the ensemble mean retains that are generally beneficial to the model and the prediction. It also identifies the properties that the original system does not have, which confuse the original solutions and are unfavorable for the prediction.

In a related study, we proved that the general concept of Lyaponov exponents for the ensemble mean system is the same as the original system. Moreover, the ensemble mean's solution is not convergent to the real solution as $t \to \infty$ and, in addition, the probability density function also changed between the ensemble mean system and the original one [15].

**Acknowledgements:** This research was jointly supported by the National Natural Sciences Foundation of China (41375112) and the National Basic Research Program of China (2011CB309704).